\documentclass[preprint, prd, aps, showpacs]{revtex4-1}

\usepackage{graphicx}
\usepackage{amsmath,amssymb}
\usepackage[usenames,dvipsnames]{color}
\usepackage[colorlinks=true, citecolor=blue, linkcolor=WildStrawberry]{hyperref}

\newcommand{\e}{\mathrm{e}} 

\newcommand{\irm}{\mathrm{i}} 
\newcommand{\beq}{\begin{equation}}
\newcommand{\eeq}{\end{equation}}
\newcommand{\bdm}{\begin{displaymath}}
\newcommand{\edm}{\end{displaymath}}

\graphicspath{{./plots/}}
\begin{document}

\title{Newtonian-noise cancellation in full-tensor gravitational-wave detectors}

\author{Jan Harms}
\affiliation{Universit\`a degli studi di Urbino ``Carlo Bo''}
\affiliation{INFN, Sezione di Firenze, Sesto Fiorentino, I-50019, Italy}
\author{Ho Jung Paik}
\affiliation{Department of Physics and Astronomy, University of Maryland, College Park, Maryland 20742}

\begin{abstract}
Terrestrial gravity noise, also known as Newtonian noise, produced by ambient seismic and infrasound fields will pose one of the main sensitivity limitations in low-frequency, ground-based, gravitational-wave (GW) detectors. It was estimated that this noise foreground needs to be suppressed by about 3 -- 5 orders of magnitude in the frequency band 10\,mHz to 1\,Hz, which will be extremely challenging. In this article, we present a new approach that greatly facilitates cancellation of gravity noise in full-tensor GW detectors. The method uses optimal combinations of tensor channels and environmental sensors such as seismometers and microphones to reduce gravity noise. It makes explicit use of the direction of propagation of a GW, and can therefore either be implemented in directional searches for GWs or in observations of known sources. We show that suppression of the Newtonian-noise foreground is greatly facilitated using the extra strain channels in full-tensor GW detectors. Only a modest number of auxiliary, high-sensitivity environmental sensors are required to achieve noise suppression by a few orders of magnitude.
\end{abstract}

\maketitle


\section{Introduction}
The advanced generation of large-scale, laser-interferometric GW detectors LIGO \cite{LSC2015} and Virgo \cite{AcEA2015} are expected to make the first direct detections of GWs within the next few years, which will open a new observational window to the Universe. The Japanese GW detector KAGRA is currently under construction and will join the detector network near the beginning of the next decade \cite{AsEA2013}. These kilometer-scale detectors are designed to observe GWs in the frequency band between 10\,Hz and a few 1000\,Hz. Upgrades of these detectors can potentially extend the band to lower frequencies by a few Hz \cite{Hil2012,Adh2013}, but today it seems infeasible to continue developing the existing facilities into detectors sensitive well below 10\,Hz. Projecting the state-of-the-art GW detector technology into the near future implemented in a detector with 10\,km arm lengths, and assuming a new detector site favorable in terms of ambient seismic noise (and associated gravity noise), it seems feasible to extend the detection band down to frequencies around 3\,Hz, as was the result of a design study for the European third-generation detector Einstein Telescope \cite{PuEA2010}. Completely new detector designs need to be considered to realize ground-based GW detectors at even lower frequencies \cite{HaEA2013}. These include the atom-interferometric \cite{DiEA2008}, the torsion-bar \cite{AnEA2010b}, and the superconducting \cite{MPC2002} GW detector concepts targeting signals between 10\,mHz and 1\,Hz.

The low-frequency sensitivity goals set for any of the potential future ground-based detectors is strongly influenced by estimates of Newtonian noise (NN). If an ideal site is selected, which means that seismic and infrasound noise are near their global minima \cite{BrEA2014}, then GW strain sensitivities of a few times $10^{-24}$\,Hz$^{-1/2}$ can be reached down to a few Hz without further gravity-noise mitigation techniques. At less favorable sites, such as the existing detector sites, or considering lower-frequency detectors, gravity-noise mitigation is required. Proposed strategies can be divided into two categories; passive and active noise mitigation. Passive mitigation aims at suppressing sources of gravity perturbation close to a detector's test masses. The detector buildings hosting the test masses act as a shield against environmental infrasound suppressing associated NN \cite{Cre2008}. The construction of moats has been proposed reflecting incoming seismic surface waves as a means to reduce NN at the LIGO sites \cite{HuTh1998}. Recess structures around the test masses can also reduce seismically induced NN \cite{HaHi2014}. However, as was explained in \cite{HaHi2014}, these techniques are effective only at higher frequencies around 10\,Hz, where the mitigating structures can have dimensions similar to the lengths of infrasound or seismic surface waves. Site selection is also considered a passive mitigation strategy. Building a GW detector underground, such as the KAGRA detector, greatly suppresses NN from seismic surface waves above a few Hz \cite{BeEA2010}. 

Whenever the passive strategies are not an option or resulting noise suppression is insufficient, active noise mitigation needs to be considered. Common to all active mitigation strategies is the usage of an array of environmental sensors with the purpose to obtain information about density perturbations in the vicinity of the test masses. Implementations of these methods then differ in how one makes use of these data. One could actively cancel the density perturbations near test masses using optimal feedback control. For example, microphones can be controlled to produce sound that cancels the ambient sound field inside a chosen volume \cite{Zan1994}. A similar scheme may be possible for seismic fields. However, this approach cannot be effective at very low frequencies where active cancellation of density fluctuations must be exerted over large volumes around test masses. Another idea is feed-forward subtraction where an estimate of the NN obtained from environmental data is used to cancel the gravity-induced motion of a test mass, or similarly, the estimate can be subtracted from the detector's data in a post-processing step. This method was first investigated in detail for the case of stationary NN using Wiener filters \cite{Cel2000}, and later also tested in numerical simulations of non-stationary seismic fields \cite{DHA2012}. While in the last publication suppression of NN from seismic surface waves was achieved using a relatively small array of about 10 seismometers, it is to be expected that especially subtraction of infrasound NN at frequencies where it is relevant (below a few Hz) requires a large number of auxiliary sensors \cite{HaEA2013}. Suppression of infrasound NN below 1\,Hz by orders of magnitude is considered an extreme challenge and potential show-stopper for low-frequency GW detectors. 

In this paper, we outline a crucial advantage of full-tensor GW detectors over conventional detectors that only output one component or combination of components of the gravity strain tensor. Full-tensor detectors measure all 5 independent components of the gravity-strain tensor as explained in Section \ref{sec:tensorGWD} for the example of superconducting GW detectors. The basic idea behind the new cancellation scheme is that a suitable linear combination of some tensor channels should make it possible to cancel NN in the remaining tensor channels. This problem is studied analytically in Section \ref{sec:math}. It will be shown that tensor NN cancellation still requires auxiliary environmental sensors, but the problem is greatly facilitated by including tensor channels. The optimal combination of tensor channels depends on the direction of propagation of a GW, and therefore the method can be applied in directional searches of GWs or observations of known GW sources. The role of sensor noise is emphasized in Section \ref{sec:Wiener}, where Wiener filters instead of the analytical expressions are introduced to find optimal channel combinations. In Section \ref{sec:discuss}, we propose a practical implementation of the method based on the simulated noise suppression using Wiener filters.

\section{Full-tensor GW detectors}
\label{sec:tensorGWD}
According to general relativity, a gravitational field is characterized by a curvature tensor.  Terrestrial laser-interferometer GW detectors measure only one off-diagonal component by combining two orthogonal light cavities.  A full-tensor detector could be constructed by measuring five degenerate quadrupole modes of a solid sphere \cite{WaPa1976,JoMe1993}.  The bandwidth of the detector could be widened by using a "split sphere," in which six test masses are suspended from a central mass \cite{PHS1996}, or a "dual sphere," in which a spherical shell encloses an inner sphere \cite{CeEA2001}.  A tensor detector is equally sensitive to GWs coming from any direction with any polarization and is thus capable of resolving the source direction and polarization.

\begin{figure}[t]
\centerline{\includegraphics[width=0.7\textwidth]{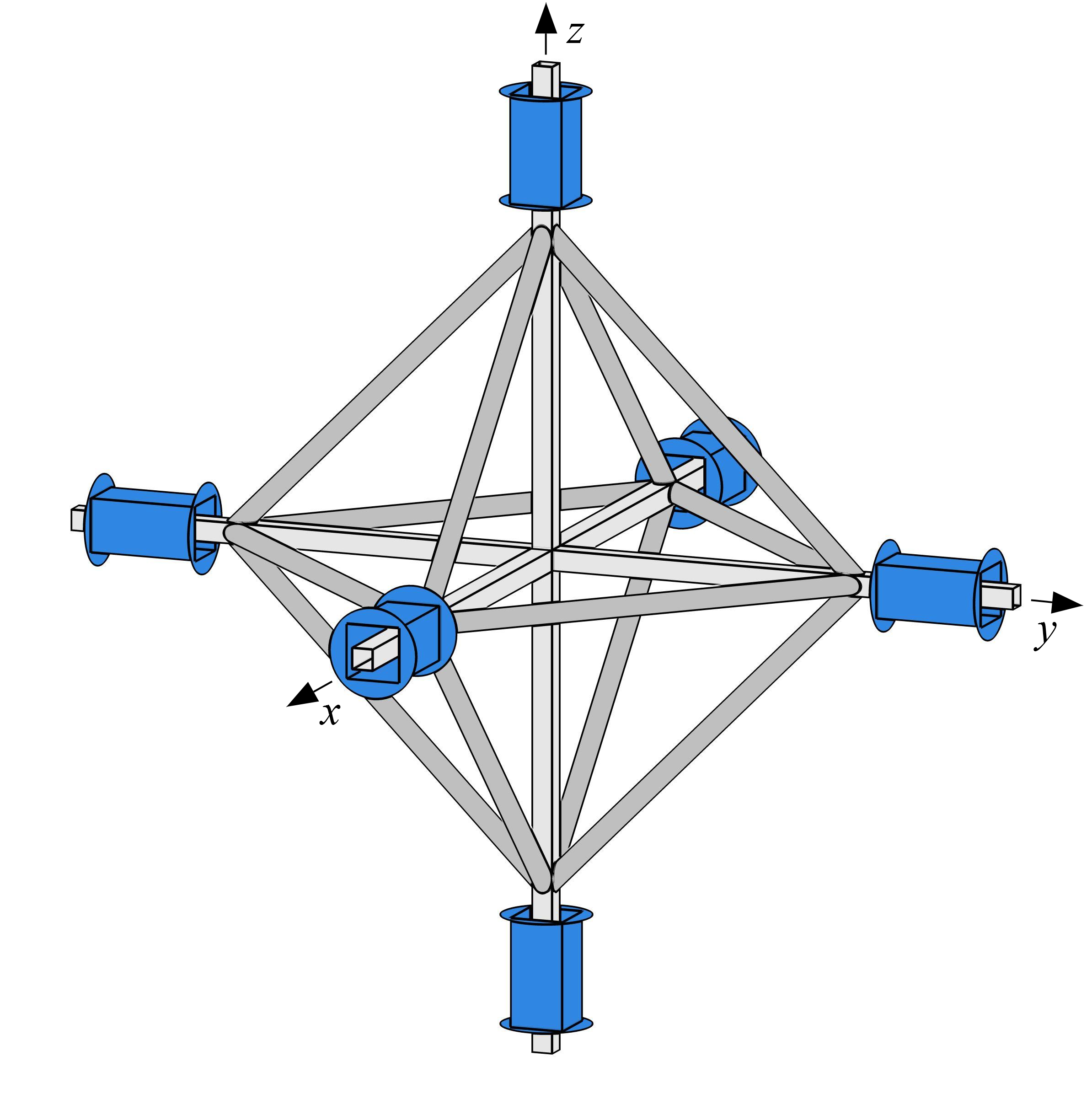}}
\caption{Test mass configuration for the low-frequency superconducting tensor GW detector. Motions of  six magnetically levitated test masses are combined to measure all six components of the curvature tensor.}
\label{fig:SOGRA}
\end{figure}
One could construct a low-frequency (0.01\,Hz to 10\,Hz) tensor GW detector by using six "almost free" test masses \cite{MPC2002}.  Figure 1 shows the test mass configuration of such a detector.  Six superconducting test masses, each with three linear degrees of freedom, are levitated over three orthogonal mounting tubes.  The test masses are made of niobium (Nb) in the shape of a rectangular shell.  Superconducting levitation/alignment coils and sensing capacitors (not shown) are located in the gap between the test masses and the mounting tubes, as well as on the outer surfaces of the test masses.  The along-axis motions of the two test masses on each coordinate axis are differenced to measure a diagonal component of the wave:
\beq
h_{ii}(\omega) = \frac{1}{L}(x_{+ii}(\omega)-x_{-ii}(\omega)),
\eeq
where $x_{\pm ij}(\omega)$ is the displacement amplitude of the test mass on the $\pm i$ axis along the $j$-th axis and $L$ is the separation between the test masses on each axis.  The cross-axis (rotational) motions of the four test masses on each coordinate plane are differenced to measure an off-diagonal component of the wave:
\beq
\begin{split}
h_{ij}(\omega) &= \frac{1}{L}\big[(x_{+ij}(\omega)-x_{-ij}(\omega))\\
&\qquad-(x_{-ji}(\omega)-x_{+ji}(\omega))\big],\,i\neq j.
\end{split}
\eeq
In addition to measuring the six strain signals, the detector will measure the three linear and three angular platform acceleration (plus gravity) signals by summing the along-axis and cross-axis test mass motions:
\beq
\begin{split}
a_{ii}(\omega) &= -\frac{\omega^2}{2}(x_{+ii}(\omega)+x_{-ii}(\omega))\\
a_{ij}(\omega) &= -\frac{\omega^2}{2}\big[(x_{+ij}(\omega)-x_{-ij}(\omega))\\
&\qquad+(x_{-ji}(\omega)-x_{+ji}(\omega))\big],\,i\neq j.
\end{split}
\eeq
These common-mode (CM) acceleration signals are used to remove the residual sensitivity of the GW detector to the platform accelerations \cite{MPC2002}.

Since test mass motion is measured with respect to the sensing circuit elements mounted on the platform, this detector requires a rigid platform with mode frequencies above the signal bandwidth.  To reduce its thermal noise, the platform itself may need to be cooled to 77\,K or lower.  To alleviate excessive demand on cryogenics, the platform must not be too heavy while it is rigid enough, with all the resonance frequencies above 10\,Hz.  The design details of this low-frequency tensor detector, called SOGRO (Superconducting Omni-directional Gravitational Radiation Observatory), will be published elsewhere (Paik et al., in preparation).

\section{Newtonian noise from infrasound and seismic surface fields}
\label{sec:math}
In this section, we present the analytical relations between NN contributions to different channels of the full-tensor GW detector. We consider the two cases of NN from seismic surface waves and infrasound, which are considered the dominant contributions to terrestrial gravity noise below 1\,Hz. Rayleigh waves are the only surface waves producing gravity perturbations. The perturbation of the gravity potential above the surface produced by Rayleigh waves is given by \cite{HaEA2013}
\beq
\delta\phi_{\rm Rf}(\vec\varrho,z,\omega)=-2\pi \frac{\gamma G\rho_0}{k} \xi(\omega)\e^{-zk}\e^{\irm\vec k\cdot\vec \varrho}.
\eeq
Here, $\vec \varrho=(x,y)$, $\vec k=k(\cos(\alpha),\sin(\alpha))$, $G$ denotes the gravitational constant, $\rho_0$ the mean mass density of the ground, $\xi(\omega)$ the amplitude of vertical surface displacement, and $\gamma\approx 0.8$ a numerical factor characteristic for fundamental Rayleigh waves that depends on the ground's Poisson's ratio. We will assume here that the horizontal wavenumber $k$ obeys the linear dispersion relation $k=\omega/c_{\rm Rf}$, but this is not important for the method and only simplifies the equations. While the perturbed gravity potential underground has additional contributions, for example, from the displacement of cavity walls, final results presented in this section are independent of this as long as the depth of the GW detector is not a significant fraction of the Rayleigh-wave length. So the choice of considering a surface detector for the Rayleigh NN calculation is just to simplify some equations. 

The gravity gradient tensor is defined as
\beq
\mathbf g(\vec r,\omega)\equiv-\nabla\otimes\nabla\delta\phi_{\rm Rf}(\vec r,\omega).
\label{eq:gravgrad}
\eeq
The gravity-gradient tensor can be identified with the second time derivative of gravity strain, $\mathbf g=\ddot{\mathbf h}$. This equivalence holds at low frequencies and for ground-based detectors where the effect of a GW can effectively be described as a tidal force acting on test masses. The response of low-frequency GW detectors to NN is described by this tensor since the distance between test masses is much smaller than the length scale of variations in the gravity field. In other words, the expression for strain NN in large-scale GW detectors,  $-\nabla\delta\phi_{\rm Rf}/L $, with $L$ being the distance between test masses, is approximately given by the second spatial derivative in Equation (\ref{eq:gravgrad}). Consequently, NN in low-frequency detectors is independent of $L$. 

For Rayleigh waves, we have
\beq
\begin{split}
\mathbf g_{\rm Rf}(\vec r&=\vec 0,\omega;\alpha)=2\pi\gamma G\rho_0 \xi(\omega)k\\
&\cdot\left(
\begin{matrix}
\cos^2(\alpha) & \cos(\alpha)\sin(\alpha) & -\irm\cos(\alpha)\\
\cos(\alpha)\sin(\alpha) & \sin^2(\alpha) & -\irm\sin(\alpha)\\
-\irm\cos(\alpha) & -\irm\sin(\alpha) & -1
\end{matrix}
\right).
\end{split}
\label{eq:gradgRf}
\eeq
An arbitrary Rayleigh-wave field can be written as a sum over many individual waves. Parameterizing the direction of propagation of a GW by angular spherical coordinates $\theta,\,\phi$, we can define a rotation $\mathbf R(\theta)\cdot\mathbf R(\phi)$ that aligns the coordinate system of the gravity-noise strain tensor $\mathbf h_{\rm Rf}$ with the propagation frame of the GW. In this case, the contribution of the GW to the total strain tensor $\mathbf h$ assumes the simple Cartesian form
\beq
\mathbf h^\prime_{\rm GW} = 
\left(\begin{matrix}
h_+ & h_\times & 0 \\
h_\times & -h_+ & 0 \\
0 & 0 & 0
\end{matrix}\right).
\eeq
In the GW propagation frame, the $z$-axis corresponds to the direction of propagation. The two rotation matrices are given by
\beq
\begin{split}
\mathbf R(\theta) &= 
\left(\begin{matrix}
\cos(\theta) & 0 & -\sin(\theta) \\
0 & 1 & 0 \\
\sin(\theta) & 0 & \cos(\theta)
\end{matrix}\right), \\
\mathbf R(\phi) &= 
\left(\begin{matrix}
\cos(\phi) & \sin(\phi) & 0 \\
-\sin(\phi) & \cos(\phi) & 0 \\
0 & 0 & 1
\end{matrix}\right),
\end{split}
\eeq
and the transformation of the gravity-noise tensor reads
\beq
\mathbf h^\prime_{\rm Rf}=\mathbf R(\theta)\cdot\mathbf R(\phi)\cdot\mathbf h_{\rm Rf}\cdot\mathbf R(-\phi)\cdot\mathbf R(-\theta).
\eeq 

Let us now take the sum $\mathbf h^\prime$ of a single GW $\mathbf h^\prime_{\rm GW}$ and Rayleigh-wave NN $\sum_i\mathbf h^\prime_{\rm Rf}(\alpha_i,\xi_i)$. It can be shown that the following relations hold:
\beq
\begin{split}
h_+=h^\prime_{11}&-2\cot(\theta) h^\prime_{13}+\cot^2(\theta)h^\prime_{33}\\
&+\csc^2(\theta)2\pi\gamma G\rho_0\frac{k}{\omega^2}\sum\limits_i\xi_i(\omega),\\
h_\times=h^\prime_{12}&-\cot(\theta) h^\prime_{23}\\
&+\irm\csc(\theta)2\pi\gamma G\rho_0\frac{k}{\omega^2}\sum\limits_i\xi_i(\omega)\sin(\alpha_i-\phi).
\end{split}
\label{eq:cancelRf}
\eeq
The sum over displacement amplitudes in the first equation simply denotes the total amplitude of vertical seismic surface displacement at the GW detector. Applying a trigonometric addition theorem, the sum over Rayleigh waves in the second equation can be rewritten in terms of horizontal seismic displacement of the Rayleigh-wave field along the two directions $x,\,y$. Therefore, a linear combination of tensor channels and one or two seismic channels (CM accelerations) can be found that perfectly cancels NN in the two target channels $h_{11},\,h_{12}$. Since the linear combination involves $\csc(\theta)$ and $\cot(\theta)$ functions that can become very large, it should be intuitively clear that the analytical relation cannot be used in practice when channels are also contaminated by additional instrumental noise. These noise contributions would be amplified by the gravity-noise cancellation. A practical solution to this problem is investigated in Section \ref{sec:Wiener}. 

Next, we repeat the calculation for infrasound NN. Here, we choose to calculate the gravity perturbation underground. We simply want to avoid the technical problem of placing the GW detector inside the fluctuating density field (i.~e.~the infrasound field), which leads to additional terms in the NN. Avoiding these terms does not change the final results or applicability of the method. Below surface, the perturbation of the gravity potential by a plane infrasound wave reflected from the surface reads
\beq
\delta\phi_{\rm IS}(\vec\varrho,z,\omega)=-4\pi\frac{G\rho_0}{\gamma p_0} \frac{\delta p(\omega)}{k^2}\e^{zk_{\rm h}}\e^{\irm\vec k_{\rm h}\cdot\vec \varrho},
\eeq
where $\delta p(\omega)$ is the amplitude of pressure fluctuations, $\gamma$ the adiabatic coefficient of air, $p_0$ the mean air pressure, $\rho_0$ the mean air mass density, and $k_{\rm h}$ the horizontal wavenumber of an infrasound wave. This leads to the gravity-gradient tensor
\beq
\begin{split}
\mathbf g_{\rm IS}(\vec r&=\vec 0,\omega;\alpha,\beta)=-4\pi\frac{G\rho_0}{\gamma p_0}\delta p(\omega)\sin^2(\beta)\\
&\cdot\left(
\begin{matrix}
\cos^2(\alpha) & \cos(\alpha)\sin(\alpha) & -\irm\cos(\alpha)\\
\cos(\alpha)\sin(\alpha) & \sin^2(\alpha) & -\irm\sin(\alpha)\\
-\irm\cos(\alpha) & -\irm\sin(\alpha) & -1
\end{matrix}
\right).
\end{split}
\eeq
The matrix is identical to the Rayleigh-wave matrix in Equation (\ref{eq:gradgRf}). As before, the angle $\alpha$ specifies the direction of propagation of the wave along the horizontal direction. The factor $\sin^2(\beta)$ is owed to the fact that the horizontal wavenumber of an infrasound wave, $k_{\rm h}=k\sin(\beta)$, depends on the angle of incidence $\beta$ with respect to the surface normal. 

The equations for the noise cancellation are given by
\beq
\begin{split}
h_+ &= h^\prime_{11}-2\cot(\theta) h^\prime_{13}+\cot^2(\theta)h^\prime_{33}\\
&\quad+\csc^2(\theta)\frac{4\pi}{\omega^2}\frac{G\rho_0}{\gamma p_0}\sum\limits_i\delta p_i(\omega)\sin^2(\beta_i),\\
h_\times &= h^\prime_{12}-\cot(\theta) h^\prime_{23}\\
&\quad+\irm\csc(\theta)\frac{4\pi}{\omega^2}\frac{G\rho_0}{\gamma p_0}\sum\limits_i\delta p_i(\omega)\sin^2(\beta_i)\sin(\alpha_i-\phi).
\end{split}
\label{eq:cancelIS}
\eeq
Here, we can see that the case of infrasound cancellation is more challenging. The sums over infrasound waves do not correspond to easily observable quantities. For example, a microphone collocated with the GW detector would observe $\sum_i\delta p_i(\omega)$ independent of the angles $\alpha_i,\,\beta_i$. Directional information could come from a gravimeter sensing associated fluctuations of the gravity field. Still, the sums cannot be rewritten in terms of gravimeter channels (CM acceleration) due to additional factors $\sin(\beta_i)$. Another problem of gravimeter channels is that they would be dominated by seismic noise at frequencies above 10\,mHz, which makes these channels useless for the cancellation of infrasound NN.

In this section, we have presented analytical expressions describing a new approach to cancel NN in full-tensor GW detectors. Cancellation of Rayleigh NN as shown in Equation (\ref{eq:cancelRf}) can be achieved with tensor channels and an additional 3-axis seismometer. In the tensor detector described in Section \ref{sec:tensorGWD}, the CM acceleration channels of the detector provide a three-axis linear and three-axis angular seismometer with SNR in excess of $10^5$ at 0.1 -- 0.3\,Hz. However, we have seen in Equation (\ref{eq:cancelIS}) that cancellation of infrasound NN is more challenging. A term remains that cannot be observed by a single microphone. Nonetheless, it is shown in Section \ref{sec:Wiener} that tensor channels greatly simplify the noise cancellation. Furthermore, while the analytical expressions cannot be used when including instrumental noise, it will be shown that efficient subtraction using only a small number of environmental sensors is still possible, for Rayleigh and infrasound NN.

\section{Newtonian noise cancellation in tensor GW detectors}
\label{sec:Wiener}
\begin{figure}[t]
\centerline{\includegraphics[width=0.7\textwidth]{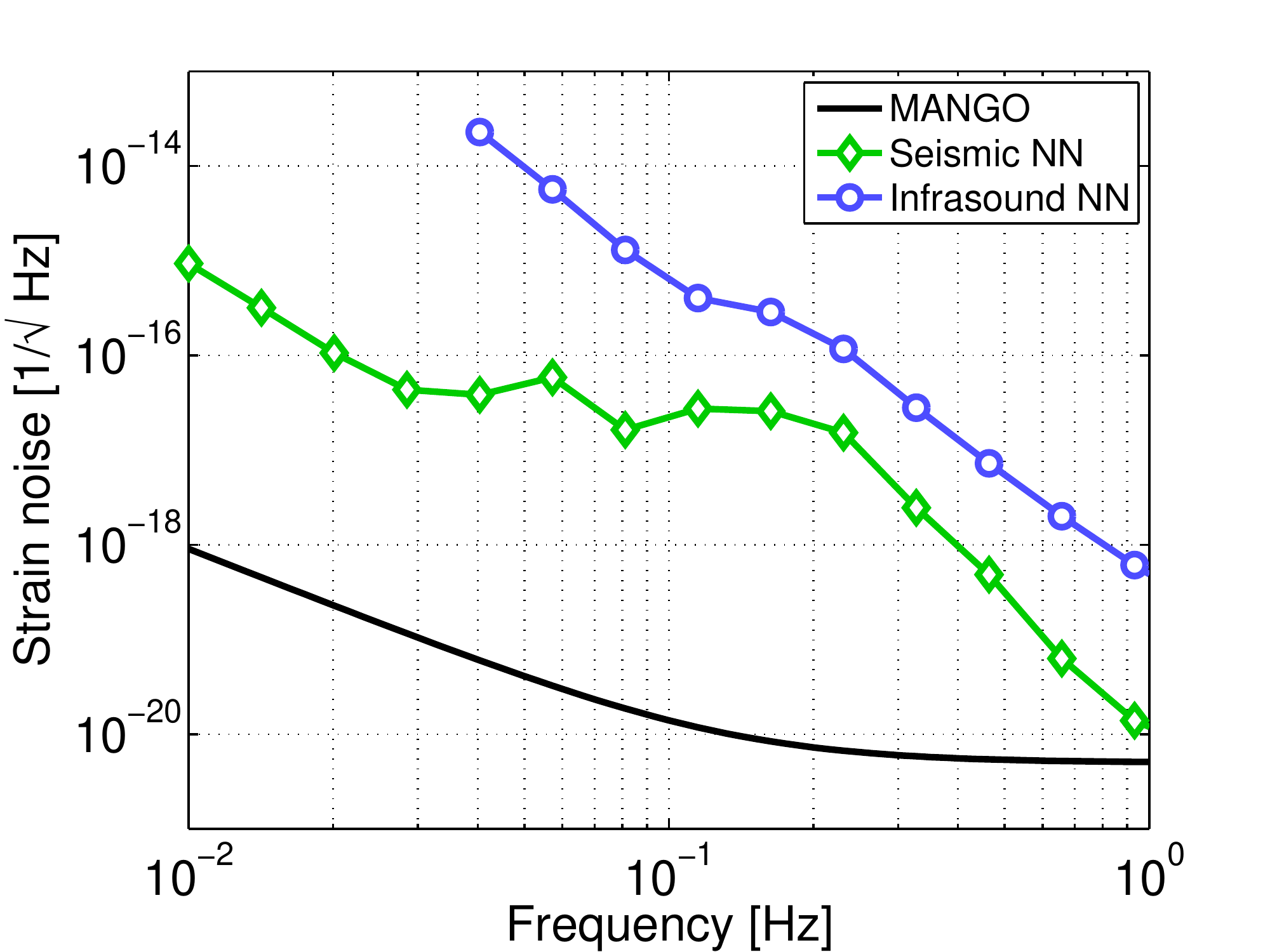}}
\caption{Sensitivity target for low-frequency GW detectors as first derived for the MANGO detector concepts \cite{HaEA2013}. The two NN estimates are based on measured seismic \cite{HaEA2010} and infrasound spectra \cite{BBB2005}.}
\label{fig:sensh}
\end{figure}

It was estimated that low-frequency GW detectors need to achieve strain sensitivities of about $10^{-20}$\,Hz$^{-1/2}$ above 0.1\,Hz in order to have good chances to observe GWs \cite{HaEA2013}. In Figure \ref{fig:sensh}, a sensitivity model is shown together with estimates of the seismic (Rayleigh) and infrasound NN. 
It can be seen that seismic NN needs to be suppressed by about 3 orders of magnitude, and infrasound NN by about 5 orders of magnitude. This is a truly daunting challenge and is rightfully considered a potential show-stopper for ground-based, low-frequency detectors. In order to achieve this suppression, it was proposed that large arrays extending over square-kilometers made of several tens to hundreds of sensors are to be deployed around the GW detector. The environmental sensors need to monitor their signals with sufficient sensitivity to avoid significant sensor-noise contributions in the cleaned strain channels. 

Assuming array configurations designed with maximized efficiency (no sensor can be removed without increasing noise residuals to an unacceptable level), the sensor signal-to-noise ratio (SNR) needs to be at least as high as the inverse of the suppression goal. For low-frequency detectors, this means that seismometers need to sense seismic displacement with SNR $>1000$, and microphones need to sense pressure fluctuations with SNR $>10^5$ at 0.1\,Hz. Requirements can be far more demanding for microphones than suggested by this rule of thumb. The additional challenge with infrasound NN cancellation is that the density perturbations are described by a 3D infrasound field, but the array can only be deployed at the surface. This greatly limits the ability to extract the required information from the infrasound measurements, and affects the optimal array configuration. Irrespective of the intrinsic sensitivity of microphones to pressure fluctuations, wind noise poses a challenge for high-sensitivity infrasound monitoring \cite{MoRa1992}. Consequently, a solution of the infrasound NN problem requires new methods and technology.

It is assumed that all noise is stationary and Gaussian, which means that optimal noise cancellation is achieved with Wiener filters \cite{BSH2008}. In frequency domain, the Wiener filter is a vector mapping reference channels $\vec R$ to an estimate $\hat n$ of the NN according to
\beq
\hat n(\omega)=\vec w(\omega)\cdot \vec R(\omega).
\eeq
This form makes use of the fact that noise at different frequencies is uncorrelated. If the Wiener filter is applied in time domain, then the last equation needs to be substituted by a convolution between the filter and the reference channels \cite{CoEA2014}. The estimated NN $\hat n$ is subsequently subtracted from the target channel. 

A Wiener filter is calculated from the matrix $C_{\rm RR}$ of correlations between reference channels, which contains the sensor noise contributions on the diagonal, and correlations $\vec C_{\rm RT}$ between reference channels and target channel. In general, correlations between channels have to be estimated from measurements, but here we assume that the density fields are isotropic (in a 2D sense for the Rayleigh field, and for the half-sphere of incident infrasound waves), which allows us to calculate the correlations precisely. Examples of calculated correlations between seismometers and gravity data can be found in \cite{BeEA2010,DHA2012}. In the notation of the previous section, the target channels of the noise cancellation are $h^\prime_{11},\,h^\prime_{12}$. The reference channels consist of all environmental sensors and the strain channels $h^\prime_{13},\,h^\prime_{23},\,h^\prime_{33}$. The components of the Wiener filter are given by
\beq
\vec w(\omega)=\vec C_{\rm RT}^\top(\omega)\cdot(C_{\rm RR}(\omega))^{-1}.
\eeq
\begin{figure}[t]
\centerline{\includegraphics[width=0.7\textwidth]{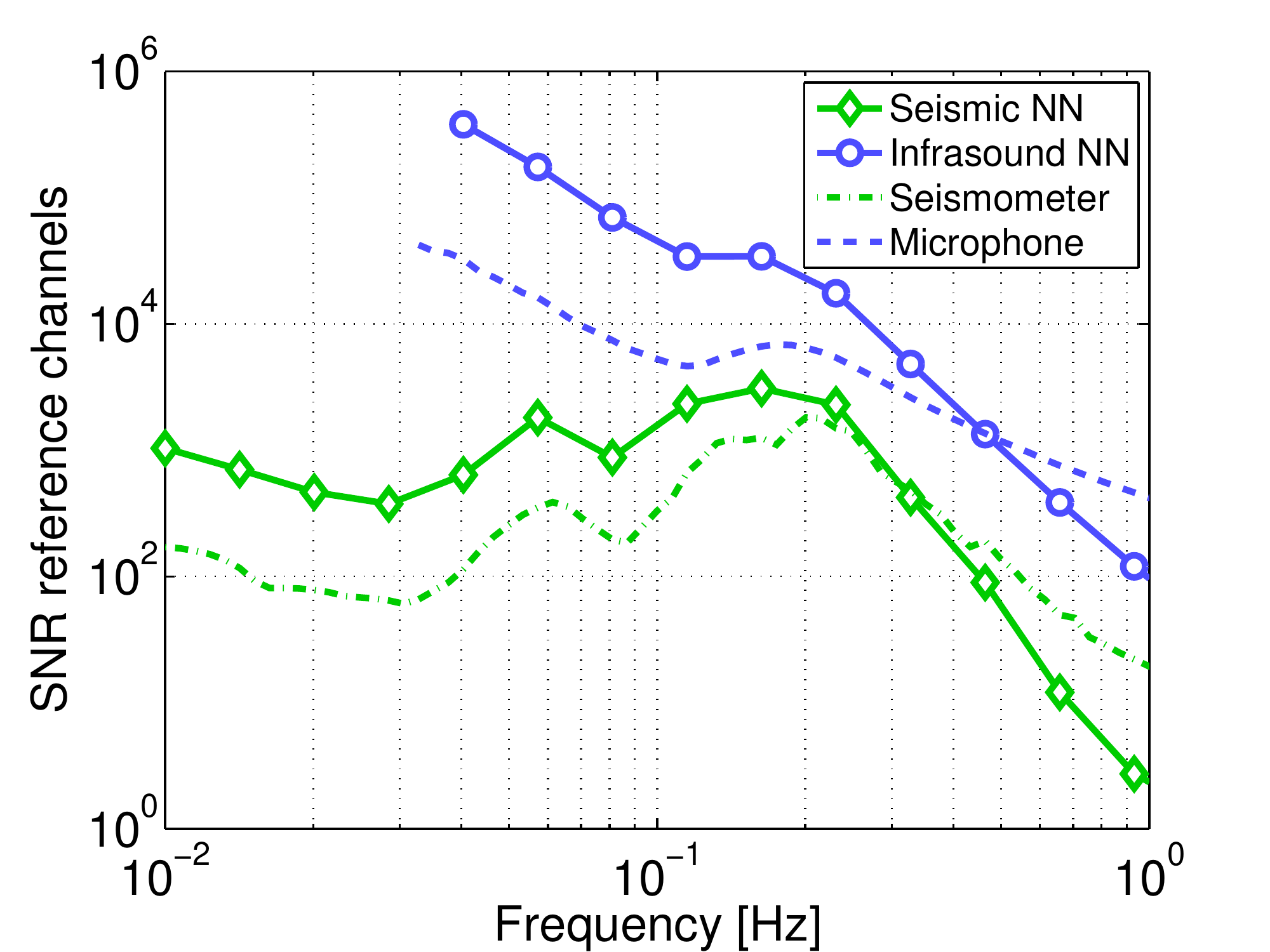}}
\caption{Estimated SNRs of reference channels. The self-noise of the seismometer lies a factor 10 below commercially available broadband instruments. The self-noise of the microphone is already achieved in current instruments. Wind noise is not included (see \cite{WaHe2009} for a recent review on wind noise reduction methods). SOGROs CM channels function as seismometers with 1000$\times$ higher SNRs.}
\label{fig:SNR}
\end{figure}
In order to evaluate the performance of a Wiener filter, we plot the residual spectrum of the target channel after NN subtraction relative to the initial spectrum $C_{\rm TT}(\omega)$ of the target channel. The residual is given by \cite{Cel2000,DHA2012}
\beq
r(\omega)=1-\frac{\vec C_{\rm RT}^\top(\omega)\cdot(C_{\rm RR}(\omega))^{-1}\cdot\vec C_{\rm RT}(\omega)}{C_{\rm TT}(\omega)}
\label{eq:resNN}
\eeq
The relative subtraction residuals $r(\omega)$ have frequency dependence since a distributed array of reference channels has frequency-dependent correlations between channels, and also since sensor SNRs vary with frequency, as shown in Figure \ref{fig:SNR}. 
\begin{figure}[t]
\centerline{\includegraphics[width=0.7\textwidth]{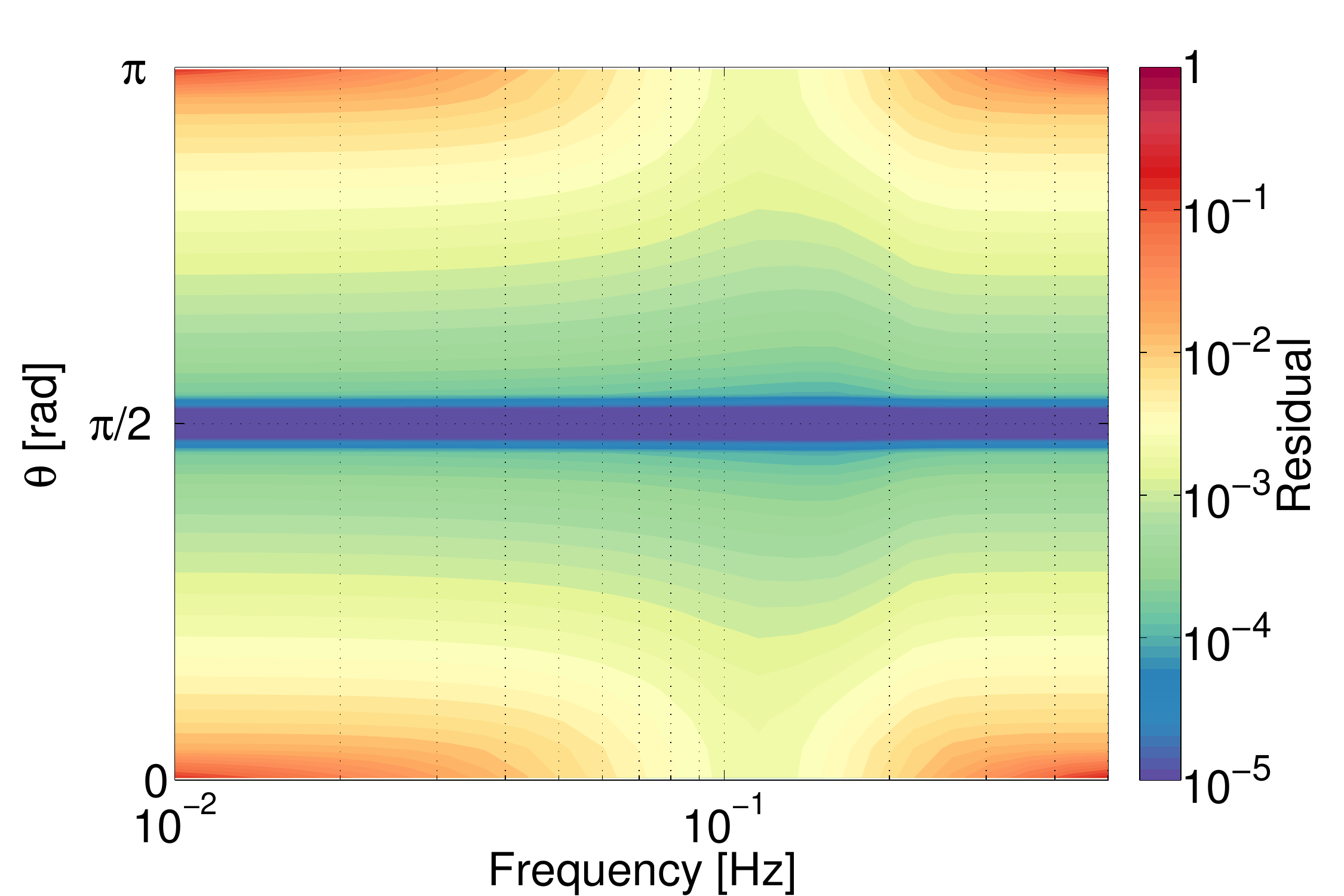}}
\caption{Relative residuals of Rayleigh NN subtraction in channel $h^\prime_{11}$ using 7 seismometers on a 5\,km ring, and the vertical CM channel. Seismometers and strain channels have $\rm SNR = 1000$. The CM channel is 1000$\times$ more sensitive than the seismometers. Residuals are independent of angle $\phi$.}
\label{fig:resRf}
\end{figure}

In the following subsections, we investigate Wiener filtering of infrasound and Rayleigh NN in detail. Results are presented only for the $h^\prime_{11}$ target channel. Noise residuals are similar for $h^\prime_{12}$, but we point out that since the Rayleigh NN cancellation ideally requires horizontal seismometer channels, there may be additional noise from surface shear waves (Love waves) that contribute to horizontal surface motion without producing gravity noise.

\subsection{Rayleigh Newtonian noise}
In the following, we demonstrate the effect of strain reference channels on residuals after subtraction of Rayleigh NN. Figure \ref{fig:resRf} shows the residual noise in $h^\prime_{11}$ using 7 seismometers on a 5\,km ring around the detector, and the vertical CM channel of the detector. The speed of Rayleigh waves is equal to 3.5\,km/s assumed here to be independent of frequency. For simplicity, we also assume that the seismometers measure seismic displacement with a \emph{frequency-independent} SNR = $1000$ (the CM channel with 1000$\times$ higher SNR), and the strain channels measure Rayleigh NN with \emph{frequency-independent} SNR = $1000$. Frequency-dependent SNRs require yet to be developed numerical tools to optimize sensor arrays used for NN cancellation. 

\begin{figure}[t]
\centerline{\includegraphics[width=0.7\textwidth]{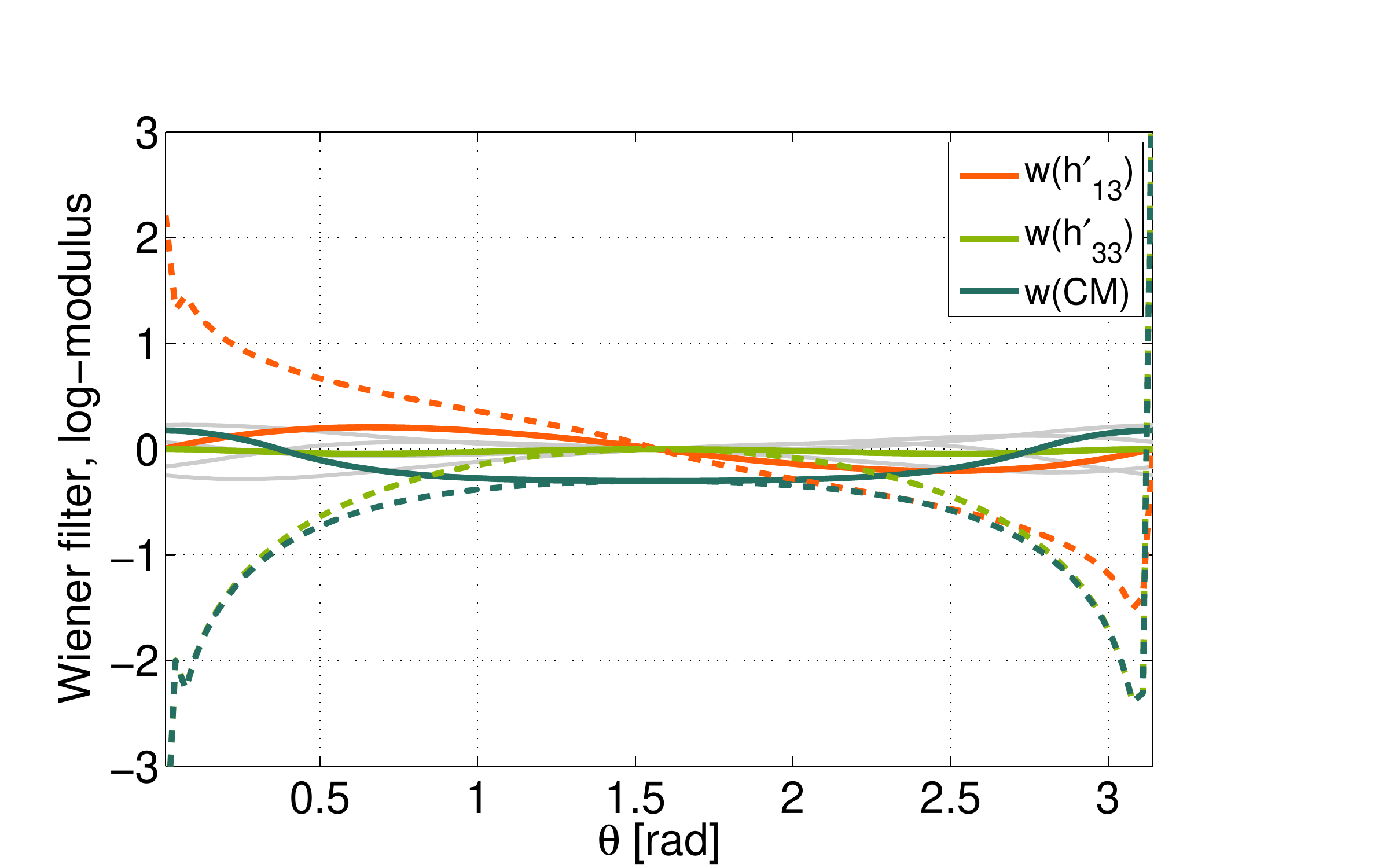}}
\caption{Filter coefficients of a noiseless Wiener filter for Rayleigh NN subtraction at 0.1\,Hz (dashed), and using reference channels with $\rm SNR = 1000$ except for the CM vertical channel, which has SNR = $10^6$ (solid). Coefficient $w(h^\prime_{23})=0$ in both cases. Gray curves in the background are for the 7 seismometers on a 5\,km ring.}
\label{fig:wienerRf}
\end{figure}

It is worth discussing in detail how this result compares to the analytical expression in Equation (\ref{eq:cancelRf}). First, if only the seismometer at the detector were used, then residuals near $\theta=0,\,\pi$ would grow to values close to 1 independent of frequency. This case is represented in Figure \ref{fig:resRf} by the residuals at frequencies $>0.3\,$Hz, where the seismometers on the ring have vanishing impact on residual noise. Residual noise close to 1 is already better than predicted by Equation (\ref{eq:cancelRf}), since the $\cot(\theta),\,\csc(\theta)$ factors mean that noise in reference channels is amplified to infinity at these angles. The Wiener filter avoids excess noise as can be understood from Figure \ref{fig:wienerRf}. It shows the non-zero Wiener filter coefficients with reference channels consisting of the strain channels, the CM vertical channel, and 7 seismometers on a 5\,km ring. Since the filter magnitude varies over orders of magnitude, the log-modulus transform, $f(x)\equiv{\rm sgn}(x)\log_{10}(1+|x|)$, was applied \cite{JoDr1980}, where $\rm sgn$ is the signum function. The dashed curves show noise residuals for infinite sensitivity reference channels. As expected, filter coefficients tend to infinity near $\theta=0,\,\pi$. The curves differ from the analytical expression in Equation (\ref{eq:cancelRf}) due to additional seismometers on the ring.
If reference channels have finite SNRs, then the Wiener filter has reduced coefficients towards the boundaries. In this way, the Wiener filter avoids injecting excess sensor noise into the target channel at the price of vanishing NN cancellation. The filter coefficients also explain why the noise residuals in Figure \ref{fig:resRf} are very small near $\theta=\pi/2$. For this value of $\theta$, the only reference used in the NN cancellation is the vertical CM channel, which has very high SNR. In comparison, the impact of the CM channel on residuals near $\theta=0,\,\pi$ is less pronounced since the residuals are dominated by noise from the strain and seismometer channels.

Figure \ref{fig:resRf} also shows that, if seismometers are added to form a ring around the detector, then subtraction performance is increased substantially for all values of $\theta$, especially within a certain frequency band. This frequency band can be chosen by adjusting the radius of the ring. The optimal radius for a certain target band depends on the Rayleigh-wave speed and on the SNR of the reference channels. The higher the SNR or the lower the speed, the smaller the optimal ring radius. 

\begin{figure}[t]
\centering{\includegraphics[width=0.7\textwidth]{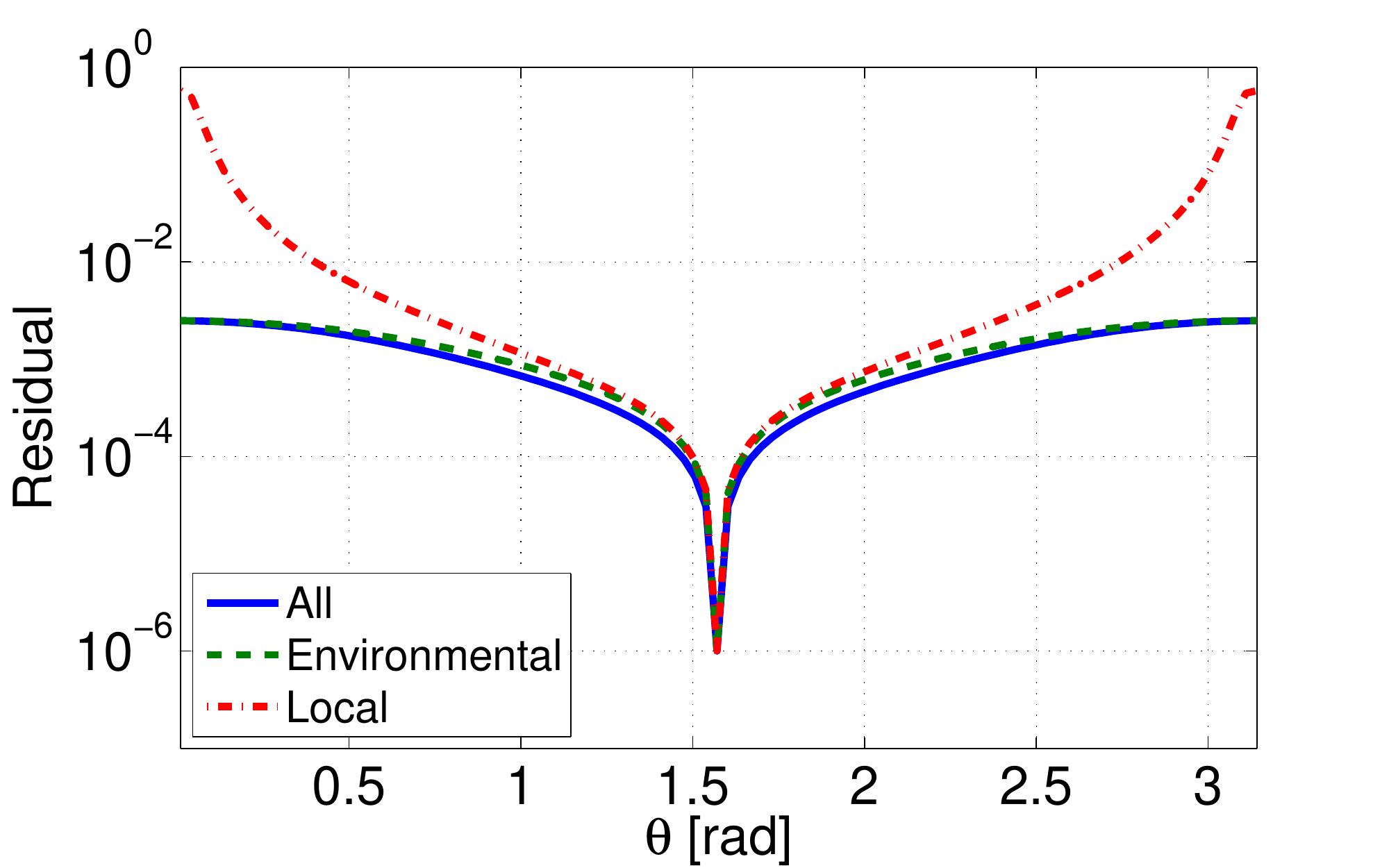}}
\caption{Relative residuals of Rayleigh NN subtraction at 0.1\,Hz in channel $h^\prime_{11}$ using vertical CM and strain channels, and 7 seismometers on a 5\,km ring. The solid curve shows the residuals including all reference channels, the dashed curve using only the 8 seismic channels, and the dotted-dashed curve using only local channels, which means the vertical CM and strain channels. Strain and seismometer channels have SNR = 1000, the CM channel SNR = $10^6$.}
\label{fig:compareRf}
\end{figure}

A direct comparison between the case with and without tensor channels is shown in Figure \ref{fig:compareRf}. As before, the configuration of the seismic array consists of 7 seismometers on a ring with 5\,km radius, the vertical CM and strain channels. It can be seen that in the case of Rayleigh NN subtraction, strain reference channels help, but do not lead to a large decrease of residuals compared to the conventional scheme based on seismometers alone. However, one should keep in mind that the residuals based on only local reference channels is competitive with the residuals of the full Wiener filter for $0.5<\theta<2.6$. Using exclusively local channels not only simplifies the experimental setup, but also potentially results in an increased robustness of the subtraction performance with respect to wave scattering and contributions from local seismic sources. These claims need to be tested in more detailed simulations.

\subsection{Infrasound Newtonian noise}
\begin{figure}[t]
\centering{\includegraphics[width=0.7\textwidth]{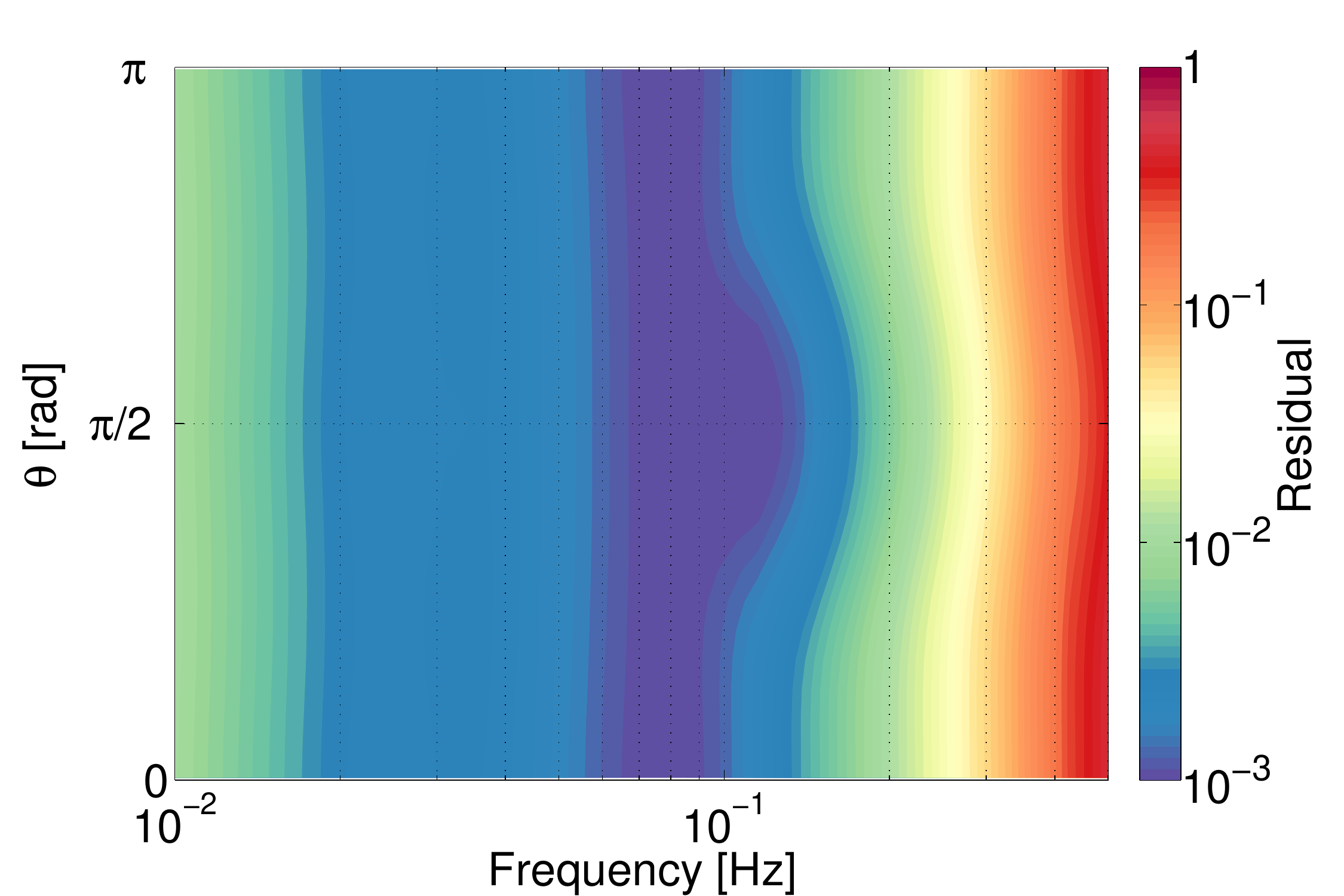}}
\caption{Relative residuals of infrasound NN subtraction in channel $h^\prime_{11}$ using 15 microphones. Microphones have $\rm SNR = 10^4$, and the strain channels $\rm SNR = 10^5$. Residuals are independent of angle $\phi$.}
\label{fig:resIS}
\end{figure}

Next, we present analogous results for infrasound NN subtraction. Noise residuals using 7 microphones on a 1\,km ring around the detector, another 7 microphones on a 600\,m ring, one microphone located at the detector, and strain channels are shown in Figure \ref{fig:resIS}. The strain channels measure infrasound NN with \emph{frequency-independent} SNR = $10^5$, and we assume that microphones measure pressure fluctuations with \emph{frequency-independent} SNR = $10^4$. With these parameter settings, noise residuals lie above the required $10^{-5}$ level. Increasing the microphone SNR and correspondingly decreasing the radii of the rings would further lower residuals, but developing such microphones will not be straight-forward. The combined effect of the two microphone rings is good broadband performance of the noise cancellation. Adding another smaller microphone ring would lower residuals at higher frequencies. While increased residuals were to be expected from Equation (\ref{eq:cancelIS}) near $\theta= 0,\,\pi$, the microphones ensure that residuals only weakly depend on $\theta$. Subtraction residuals at 0.1\,Hz would be greater by almost an order of magnitude if only a single microphone ring were deployed. 

Subtraction performance is generally worse compared to the Rayleigh NN case since infrasound NN is caused by a 3D wavefield that is monitored by a microphone array constrained to Earth surface. Being composed of evanescent waves, the Rayleigh seismic field also displaces the ground below surface, but since Rayleigh waves are genuinely surface waves, a seismic array deployed at the surface can extract all information about associated density fluctuations. 

\begin{figure}[t]
\centering
\includegraphics[width=0.7\textwidth]{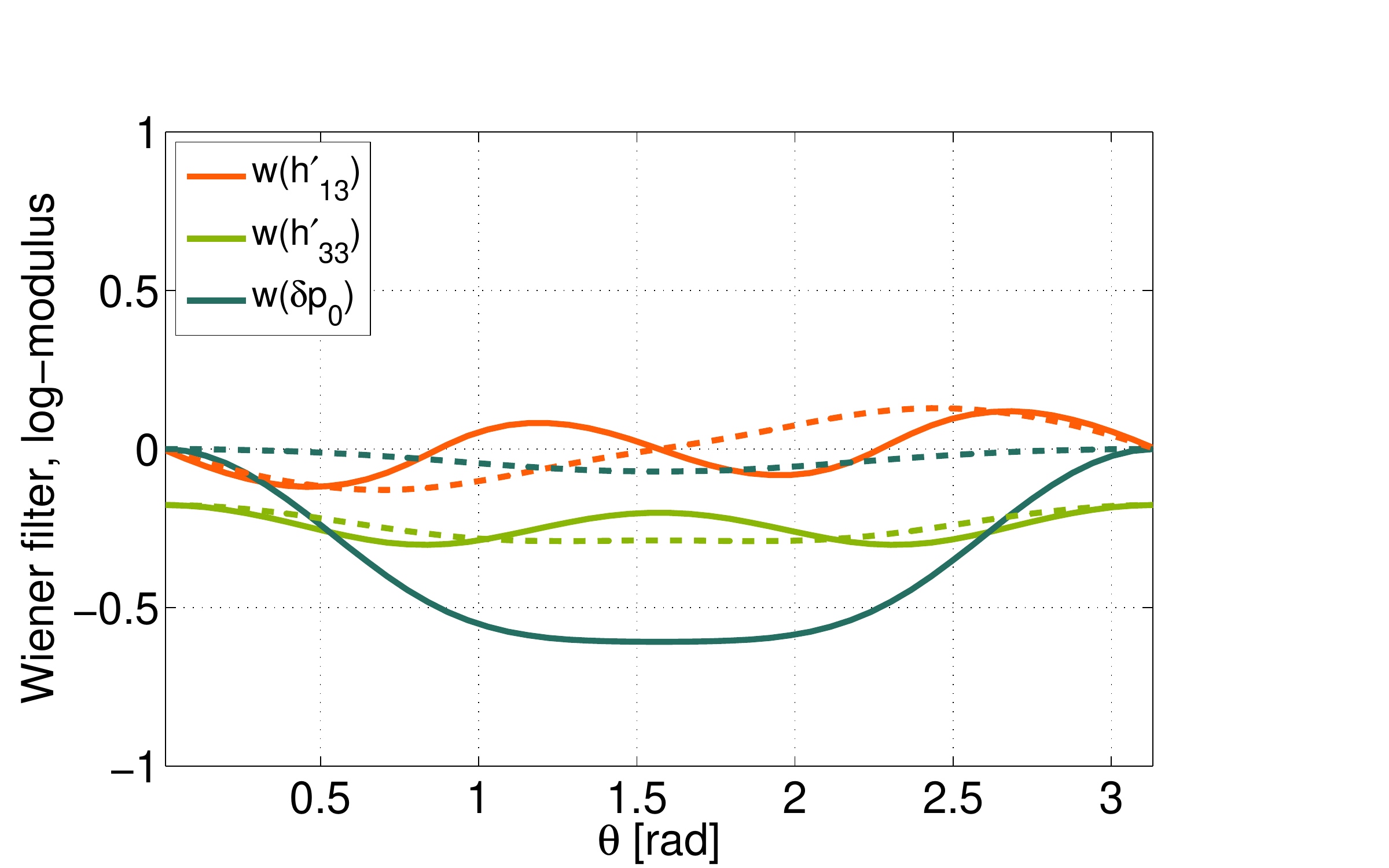}
\caption{Wiener-filter coefficients using microphones with SNR = $10^4$ and strain channels with SNR = $10^5$ for infrasound NN subtraction. Coefficient $w(h^\prime_{23})=0$ in all cases. Dashed: Single microphone located at the detector.  Solid: Seven microphones on a 600\,m ring around the detector and a single microphone located at the detector. Subtraction is calculated at 0.1\,Hz. Only the coefficient for the microphone at the detector is plotted.}
\label{fig:wienerIS}
\end{figure}

The Wiener filter coefficients in Figure \ref{fig:wienerIS} confirm that the infrasound NN subtraction is more challenging. First, the plotted filter coefficients are far off the analytical expression in Equation (\ref{eq:cancelIS}). This was to be expected since Equation (\ref{eq:cancelIS}) states that no simple combination of local reference channels can provide an accurate estimate of NN. In contrast to the Rayleigh NN case, where good subtraction can also be achieved with local channels only, at least for a range of values of $\theta$, deployment of a microphone array is essential to achieve good subtraction performance. The dashed curves show the filter coefficients with a single microphone located at the detector, while the solid curves are calculated for a Wiener filter that includes 7 additional microphones on a 600\,m ring. Microphones measure pressure fluctuations with SNR = $10^4$, and strain channels measure infrasound NN with SNR = $10^5$. Comparison between the two sets of curves in Figure \ref{fig:wienerIS} shows that the microphones on the ring contribute substantially to the noise cancellation, since the coefficients are greatly modified when including the extra microphones. Among all local channels, only $h^\prime_{33}$ contributes to a NN estimate near $\theta = 0,\,\pi$ in both cases. Coefficients of the Wiener filter for infrasound NN subtraction generally depend weakly on the SNR of reference channels. The coefficients in Figure \ref{fig:wienerIS} would look the same for any SNR values greater than 10.

In order to achieve the low residuals in Figure \ref{fig:resIS}, it was necessary to deploy two rings of microphones, a total of 15 microphones being used for the noise cancellation. At this point, one may wonder if the strain reference channels still contribute significantly to the subtraction performance. Figure \ref{fig:compareIS} shows that, in contrast to Rayleigh NN, reference strain channels play a very important role in infrasound NN subtraction. The dotted-dashed curve shows the noise residuals if only local channels are used (i.~e.~strain channels and one microphone). Subtraction performance is very poor. The dashed curve shows subtraction residuals if only microphones are used. In this case, subtraction can be fairly good, but only for certain values of $\theta$. If all channels are included, then excellent subtraction performance is achieved for all values of $\theta$ (solid curve). This means that microphone array and strain channels both play an important role. 

\section{Discussion}
\label{sec:discuss}
\begin{figure}[t]
\centering{\includegraphics[width=0.7\textwidth]{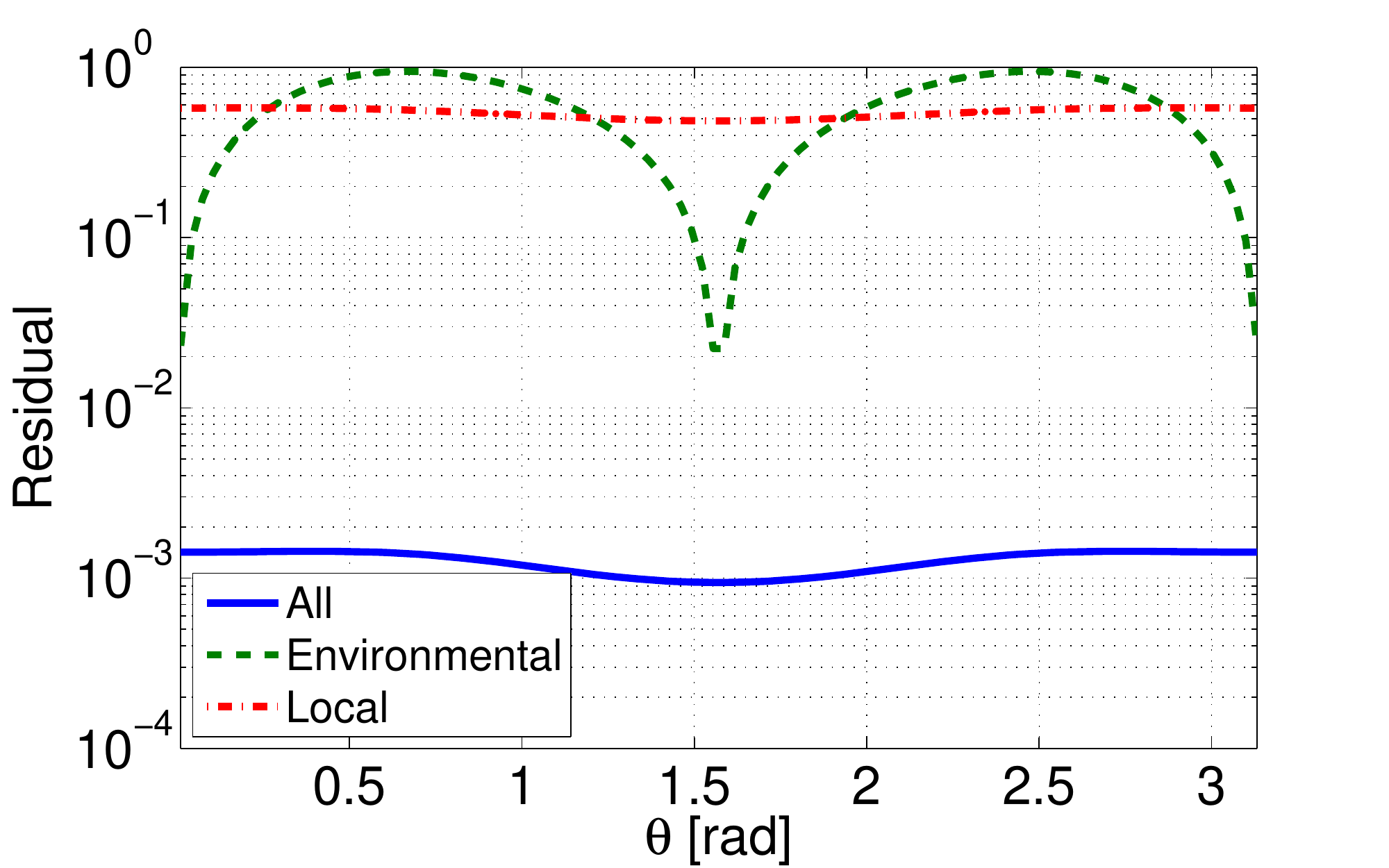}}
\caption{Relative residuals of infrasound NN subtraction at 0.1\,Hz in channel $h^\prime_{11}$ using strain channels and 15 microphones. The solid curve shows the residuals including all strain and microphone reference channels, the dashed curve using only microphones, and the dotted-dashed curve using only local channels, which means the strain channels and one microphone located at the detector. Strain channels have SNR = $10^5$ and microphone channels have SNR = $10^4$.}
\label{fig:compareIS}
\end{figure}

Combining Figures \ref{fig:compareRf} and \ref{fig:compareIS}, we come to a clear conclusion towards an efficient scheme of NN cancellation in full-tensor GW detectors. Since Wiener filter coefficients of the strain channels for Rayleigh and infrasound NN are different, it is impossible to cancel both simultaneously. However, we have seen that Rayleigh NN can be cancelled efficiently without including strain channels, whereas subtraction of infrasound NN profits greatly from strain channels. Consequently, we propose the following strategy. As a first step, Rayleigh NN needs to be cancelled in all 5 independent strain channels. Good broadband subtraction performance can be achieved with spiral seismometer arrays consisting of a few tens of seismometers with SNR = 1000 at the microseismic peak \cite{HaEA2013}. The required sensitivity of the seismometers is also near the sensitivity of available commercial instruments \cite{Rod1994}. When the strain channels are cleaned from Rayleigh NN, the next step is to subtract infrasound NN using microphones and the strain channels. We have seen that noise residuals are greatly reduced compared to subtraction with microphones only. Information about the 3D infrasound field, which cannot be retrieved with a 2D microphone array is partially provided by the strain channels. Subtraction residuals still do not reach the sensor-noise limit, but compared to the scheme with microphones only, residuals are lowered by orders of magnitude depending on the direction of propagation of the GW. Therefore, we conclude that the problem of NN mitigation in low-frequency GW detectors is greatly facilitated in full-tensor detectors. 

Certain aspects of low-frequency NN cancellation demand or deserve a deeper analysis. First of all, SNRs of the sensors are frequency-dependent although we used frequency-independent SNRs to simplify our calculation.  In a real experiment, the NN would be removed using Wiener filters with sensor arrays optimized for the actual measured SNRs. Numerical tools for array optimization need to be developed. Second, it is conceivable that mitigation schemes relying on local channels rather than on data from large distributed arrays are more robust against wave scattering and disturbance from local sources, which both influence the spatial correlation of seismic or infrasound fields. Clearly, it is always easier to optimize a noise cancellation based on local channels, but there may be additional advantages of using strain channels with respect to robustness of the performance. Robustness can play a very important role since gravity perturbations need to be understood at a level 1/1000 for Rayleigh NN or $1/10^5$ for infrasound NN. Especially for the infrasound microphones, it is unclear whether such level of accuracy in the monitoring itself can be achieved. Here, we do not refer to the intrinsic readout noise of the sensors, but environmental noise, for example, from wind \cite{MoRa1992}. 

Also, it is not yet understood if a change of spatial correlation due to scattering or local sources poses a limitation to the subtraction performance, which can only be overcome by deploying additional environmental sensors, or if it leads to modification of NN and density fields in such a way that a simple rearrangement of senors can compensate the loss in subtraction performance without increasing the number of sensors. These considerations play an important role for high subtraction goals and at low frequencies since it is highly unlikely that a location can be identified where heterogeneities of the ground and surface profile are negligible over the relevant volumes \cite{CoHa2012}. Nonetheless, all these challenges exist for any type of NN mitigation at low frequencies, and one may expect that they impact coherent noise mitigation more strongly in schemes based on large sensor arrays than on schemes partially relying on strain channels and a smaller number of environmental sensors. 

Finally, given the fact that infrasound NN needs to be understood at a level $1/10^5$ for a complete cancellation of the noise, approximations in our gravity models such as negligible size of the GW detector or placing the detector at the surface despite the fact that it may be a few hundred meters underground potentially lead to significant modelling errors. More accurate models need to be investigated to find out if the mitigation scheme proposed in this article needs to be modified for the most ambitious noise-suppression goals.

\acknowledgments
The work of JH was supported by a Marie-Curie Fellowship (FP7-PEOPLE-2013-IIF). The work of HJP was supported by NSF grant PHY1105030 and NASA grant NNX14AI43G.


\end{document}